\documentstyle[prl,aps,floats,twocolumn]{revtex}

\tighten

\input epsf.tex

\newcommand{\postscript}[2]{\setlength{\epsfxsize}{#2\hsize}
   \centerline{\epsfbox{#1}}}
\newcommand{\sign}{\:\!\text{sign}\:\!}
\newcommand{\mgaugino}{M_{1/2}}

\newcommand{\maux}{M_{\text{aux}}}
\newcommand{\msusy}{M_{\text{SUSY}}}
\newcommand{\mgut}{M_{\text{GUT}}}

\newcommand{\mlosp}{M_{\text{LOSP}}}
\newcommand{\tb}{\tan\beta}

\newcommand{\gev}{\text{GeV}}
\newcommand{\tev}{\text{TeV}}

\newcommand{\be}{\begin{equation}}
\newcommand{\ee}{\end{equation}}
\newcommand{\etal}{{\em et al.}}
\newcommand{\eg}{{\em e.g.}}

\newcommand{\bsg}{B\to X_s \gamma}

\newcommand{\amu}{a_{\mu}}
\newcommand{\amuexp}{a_{\mu}^{\text{exp}}}
\newcommand{\amususy}{a_{\mu}^{\text{SUSY}}}
\newcommand{\amusm}{a_{\mu}^{\text{SM}}}

\newcommand{\eqref}[1]{Eq.~(\ref{#1})}

\begin{document}

\draft

\renewcommand{\thefootnote}{\fnsymbol{footnote}}
\setcounter{footnote}{0}

\preprint{
\noindent
\hfill
\begin{minipage}[t]{3in}
\begin{flushright}
MIT--CTP--3083\\
CERN--TH/2001--039\\
hep-ph/0102146
\end{flushright}
\end{minipage}
}

\twocolumn[\hsize\textwidth\columnwidth\hsize\csname
@twocolumnfalse\endcsname

\title{Supersymmetry and the Anomalous Anomalous Magnetic Moment of
the Muon }

\author{
Jonathan L.~Feng$^a$
and Konstantin T.~Matchev$^b$
\vskip 0.1in
}

\address{
  ${}^{a}$
  Center for Theoretical Physics,
  Massachusetts Institute of Technology\\
  Cambridge, MA 02139, U.S.A.\\ 
\vskip 0.1in
  ${}^{b}$
  Theory Division, CERN, CH--1211
  Geneva 23, Switzerland}


\maketitle

\begin{abstract} 
The recently reported measurement of the muon's anomalous magnetic
moment differs from the standard model prediction by $2.6\sigma$.  We
examine the implications of this discrepancy for supersymmetry.
Deviations of the reported magnitude are generic in supersymmetric
theories.  Based on the new result, we derive model-independent upper
bounds on the masses of observable supersymmetric particles.  We also
examine several model frameworks.  The sign of the reported deviation
is as predicted in many simple models, but disfavors anomaly-mediated
supersymmetry breaking.
\end{abstract}





]

\renewcommand{\thefootnote}{\arabic{footnote}}
\setcounter{footnote}{0}

Measurements of spin magnetic dipole moments have a rich history as
harbingers of profound progress in particle physics.  In the leptonic
sector, the electron's gyromagnetic ratio $g_e \approx 2$ pointed the
way toward Dirac's theory of the electron.  Later, the electron's
anomalous magnetic moment $a_e \equiv (g_e -2)/2 \approx \alpha /
2\pi$ played an important role in the development of quantum
electrodynamics and renormalization.  Since then, increasingly precise
measurements have become sensitive both to very high order effects in
quantum electrodynamics and to hadronic processes, and the consistency
of experiment and theory has stringently tested these sectors of the
standard model.

Very recently, the Muon ($g-2$) Collaboration has reported a
measurement of the muon's anomalous magnetic moment, which, for the
first time, is sensitive to contributions comparable to those of the
weak interactions~\cite{Brown:2001mg}. (See Tables~\ref{table:I} and
\ref{table:II}.)  The new Brookhaven E821 result is $\amuexp = 11\
659\ 202\, (14)\, (6) \times 10^{-10}\ (1.3\ \text{ppm})$, where the
first uncertainty is statistical and the second systematic.  Combining
experimental and theoretical uncertainties in quadrature, the new
world average differs from the standard model prediction by $2.6
\sigma$~\cite{Brown:2001mg}:
\begin{equation}
\amuexp - \amusm = (43 \pm 16) \times 10^{-10} \ .
\end{equation}

Although of unprecedented precision, the new result is based on a
well-tested method used in previous measurements.  Polarized positive
muons are circulated in a uniform magnetic field. They then decay to
positrons, which are emitted preferentially in the direction of the
muon's spin.  By analyzing the number of energetic positrons detected
at positions around the storage ring, the muon's spin precession
frequency and anomalous magnetic moment are determined.  The new
result is based solely on 1999 data.  Analysis of the 2000 data is
underway, with an expected error of $\sim 7 \times 10^{-10}$ (0.6
ppm), and the final goal is an uncertainty of $4 \times 10^{-10}$
(0.35 ppm)~\cite{Carey}.

The standard model prediction has been greatly refined in recent
years.  The current status is reviewed in Ref.~\cite{Czarnecki:2000id}
and summarized in Table~\ref{table:II}.  The uncertainty is dominated
by hadronic vacuum polarization contributions that enter at 2-loops.
This is expected to be reduced by recent measurements of
$\sigma(e^+e^- \to \text{hadrons})$ at center-of-mass energies
$\sqrt{s} \sim 1~\gev$.  Thus, although the statistical significance
of the present deviation leaves open the possibility of agreement
between experiment and the standard model, the prospects for a
definitive resolution are bright.  If the current deviation remains
after close scrutiny and the expected improvements, the anomalous
value of $\amu$ will become unambiguous.

In this study, we consider the recent measurement of $\amu$ to be a
signal of physics beyond the standard model.  In particular, we
consider its implications for supersymmetric theories.  Supersymmetry
is motivated by many independent considerations, ranging from the
gauge hierarchy problem to gauge coupling unification to the necessity
of non-baryonic dark matter, all of which require supersymmetric
particles to have weak scale masses.  Deviations in $\amu$ with the
reported magnitude are therefore generic in supersymmetry.  In
addition, $\amu$ is both flavor- and CP-conserving.  Thus, while the
impact of supersymmetry on other low energy observables can be highly
suppressed by scalar degeneracy or small CP-violating phases in simple
models, supersymmetric contributions to $\amu$ cannot be.  In this
sense, $\amu$ is a uniquely robust probe of supersymmetry, and an
anomaly in $\amu$ is a natural place for the effects of supersymmetry
to appear.

The anomalous magnetic moment of the muon is the coefficient of the
operator $\amu \frac{e}{4m_\mu} \, \bar{\mu} \sigma^{mn} \mu \,
F_{mn}$, where $\sigma^{mn} = \frac{i}{2} \left[ \gamma^m, \gamma^n
\right]$.  The supersymmetric contribution, $\amususy$, is dominated
by well-known neutralino-smuon and chargino-sneutrino
diagrams~\cite{SUGRA}.  In the absence of significant slepton flavor
violation, these diagrams are completely determined by only seven
supersymmetry parameters: $M_1$, $M_2$, $\mu$, $\tb$,
$m_{\tilde{\mu}_L}$, $m_{\tilde{\mu}_R}$, and $A_{\mu}$.  The first
four enter through the chargino and neutralino masses: $M_1$, $M_2$,
and $\mu$ are the U(1) gaugino, SU(2) gaugino, and Higgsino mass
parameters, respectively, and $\tb = \langle H^0_u \rangle / \langle
H^0_d \rangle$ governs gaugino-Higgsino mixing.  The last five
determine the slepton masses, where $m_{\tilde{\mu}_L}$ and
$m_{\tilde{\mu}_R}$ are the SU(2) doublet and singlet slepton masses,
respectively, and the combination $m_{\mu} (A_{\mu} - \mu \tb)$ mixes
left- and right-handed smuons.  In general, $M_1$, $M_2$, $\mu$, and
$A_{\mu}$ are complex.  However, bounds from electric dipole moments
typically require their phases to be very small.  In addition,
$|\amususy|$ is typically maximized for real parameters.  In deriving
model-independent upper bounds on superparticle masses below, we
assume real parameters, but consider all possible sign combinations;
these results are therefore valid for arbitrary phases.  Our sign
conventions are as in Ref.~\cite{Feng:2000hg}.

The qualitative features of the supersymmetric contributions are most
transparent in the mass insertion approximation. The structure of the
magnetic dipole moment operator requires a left-right transition along
the lepton-slepton line.  In the interaction basis, this transition
may occur through a mass insertion in an external muon line, at a
Higgsino vertex, or through a left-right mass insertion in the smuon
propagator.  The last two contributions are proportional to the muon
Yukawa coupling and so may be enhanced by $\tb$.  For large and
moderate $\tb$, it is not hard to show that the supersymmetric
contributions in the mass insertion approximation are all of the form
\begin{equation}
\frac{g_i^2}{16 \pi^2} m_{\mu}^2\, \mu M_i  \tb \, F \ ,
\label{massinsertion}
\end{equation} 
where $i=1,2$, and $F$ is a function of superparticle masses, with $F
\propto \msusy^{-4}$ in the large mass limit~\cite{Moroi:1996yh}.

Equation~(\ref{massinsertion}) implies $\amususy/a_e^{\text{SUSY}}
\sim m_{\mu}^2 / m_e^2 \approx 4 \times 10^4$; $\amu$ is therefore far
more sensitive to supersymmetric effects than $a_e$, despite the fact
that the latter is 350 times better measured.  Also, for $M_2/M_1 >
0$, although the contributions of \eqref{massinsertion} may
destructively interfere, typically $\sign(\amususy) = \sign(\mu
M_{1,2})$; we have found exceptions only rarely in highly
model-independent scans. Finally, the parameter $\tb$ is expected to
be in the range $2.5 \alt \tb \alt 50$, where the lower limit is from
Higgs boson searches, and the upper limit follows from requiring a
perturbative bottom quark Yukawa coupling up to $\sim
10^{16}~\gev$. Supersymmetric contributions may therefore be greatly
enhanced by large $\tb$.
\begin{table}[tb]
\caption{Recent measurements of $\amu \times 10^{10}$ and the
cumulative world average. 
\label{table:I} 
}
\begin{tabular}{ll@{ }lr}
Data Set
 & \multicolumn{2}{c}{Result}
  & \multicolumn{1}{c}{World Average}  \\ \hline
 CERN77~\cite{Bailey:1979mn} \rule[0mm]{0mm}{4mm}
 & $11\ 659\ 230\, (85)$ & ($7$ ppm)
  &  \multicolumn{1}{c}{\text{---}}  \\  
 BNL97~\cite{Carey:1999dd}
 & $11\ 659\ 250\, (150)$ & ($13$ ppm)
  & $11\ 659\ 235\, (73)$ \\
 BNL98~\cite{Brown:2000sj}
 & $11\ 659\ 191\, (59)$ & ($5$ ppm)
  & $11\ 659\ 205\, (46)$ \\
 BNL99~\cite{Brown:2001mg}
 & $11\ 659\ 202\, (14)\, (6)$ & ($1.3$ ppm)
  & $11\ 659\ 203\, (15)$ 
\end{tabular}
\vspace*{-.1in}
\end{table}

\begin{table}[tbh]
\caption{Contributions to the standard model prediction for $\amu
\times 10^{10}$. (See Ref.~\protect\cite{Czarnecki:2000id} and
references therein.)
\label{table:II}
}
\begin{tabular}{lr}
Standard model source
  & \multicolumn{1}{c}{\ Contribution}  \\ \hline
 QED (up to 5-loops) \rule[0mm]{0mm}{4mm}
  & $11\ 658\ 470.6\, (0.3)$\\
 Hadronic vac. polarization (2-loop piece)
  & $692.4\, (6.2)$  \\
 Hadronic vac. polarization (3-loop piece)
  & $-10.0\, (0.6)$  \\
 Hadronic light-by-light
  & $-8.5\, (2.5)$ \\
 Weak interactions (up to 2-loops)
  & $15.2\, (0.4)$  \\
 {\em Total}
  & $11\ 659\ 159.7\, (6.7)$ \\ 
\end{tabular}
\vspace*{-.1in}
\end{table}
To determine the possible values of $\amususy$ without model-dependent
biases, we have calculated $\amususy$ in a series of high statistics
scans of parameter space.  We use exact mass eigenstate expressions
for $\amususy$.  Our calculations agree with
Refs.~\cite{Moroi:1996yh,Carena:1997qa,Blazek:1999hb} and cancel the
corresponding standard model diagrams in the supersymmetric
limit~\cite{Ferrara:1974wb}.  We require chargino masses above 103 GeV
and smuon masses above 95 GeV~\cite{masslimits}.  We also assume that
the lightest supersymmetric particle (LSP) is stable, as in
gravity-mediated theories, and require it to be neutral.  Finally, for
each scan point we record the mass and identity of the lightest
observable supersymmetric particle (LOSP), which we define to be the
lightest superpartner with decay products that are detectable at
colliders. Given the assumption of a stable LSP, the LOSP is the
second lightest supersymmetric particle, or the third if the two
lightest are a neutralino and the sneutrino.  Note that the
identification of supersymmetric events is a complicated and
model-dependent issue, especially at hadron colliders. The LOSP, as
defined here, is not guaranteed to be observed above background, even
if produced.

\begin{figure}[tbp]
\postscript{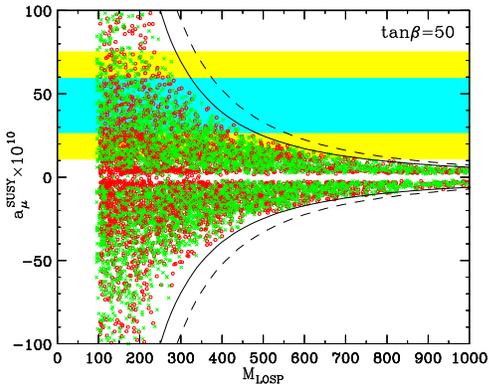}{0.74}
\caption{Allowed values of $M_{\text{LOSP}}$, the mass of the lightest
observable supersymmetric particle, and $\amususy$ from a scan of
parameter space with $M_1=M_2/2$, $A_{\mu} = 0$, and $\tb = 50$.
Green crosses (red circles) have smuons (charginos/neutralinos) as the
LOSP.  The 1$\sigma$ and 2$\sigma$ allowed $\amususy$ ranges are
indicated. Relaxing the relation $M_1=M_2/2$ leads to the solid
envelope curve, and further allowing arbitrary $A_{\mu}$ leads to the
dashed curve.  The envelope contours scale linearly with $\tb$.  A
stable LSP is assumed.  }
\label{fig:mnlsp}
\end{figure}

We begin by scanning over the parameters $M_2$, $\mu$,
$m_{\tilde{\mu}_L}$, and $m_{\tilde{\mu}_R}$, assuming gaugino mass
unification $M_1= M_2/2$, $A_{\mu}=0$, and $\tb=50$.  The free
parameters take values up to 2.5 TeV.  The resulting values in the
$(\mlosp, \amususy)$ plane are given by the points in
Fig.~\ref{fig:mnlsp}.  We then consider arbitrary (positive and
negative) values of $M_2/M_1$.  The resulting values are bounded by
the solid curve.  As can be seen, and as verified by high statistics
sampling targeting the border area, the assumption of gaugino mass
unification has no appreciable impact on the envelope curve. Finally,
we allow any $A_{\mu}$ in the interval $[-100~\tev, 100~\tev]$.  The
resulting sample is extremely model-independent, and is bounded by the
dashed contour of Fig.~\ref{fig:mnlsp}. The envelope contours scale
linearly with $\tb$ to excellent approximation.

{}From Fig.~\ref{fig:mnlsp} we see that the measured deviation in
$\amu$ is in the range accessible to supersymmetric theories and is
easily explained by supersymmetric effects.  

The anomaly in $\amu$ also has strong implications for the
superpartner spectrum.  Among the most important is that at least two
superpartners cannot decouple if supersymmetry is to explain the
deviation, and one of these must be charged and so observable at
colliders.  Non-vanishing $\amususy$ thus imply upper bounds on
$\mlosp$.  The large value of $\tb$ is chosen to allow the largest
possible $\mlosp$.  The solid contour is parametrized by
\begin{equation}
\frac{\amususy}{43\times 10^{-10}} = \frac{\tb}{50} \left(
\frac{390~\gev}{\mlosp^{\text{max}}} \right)^2 \ .
\end{equation}
If $\amususy$ is required to be within 1$\sigma$ (2$\sigma$) of the
measured deviation, at least one observable superpartner must be
lighter than 490 GeV (800 GeV).  

\begin{figure}[tbp]
\postscript{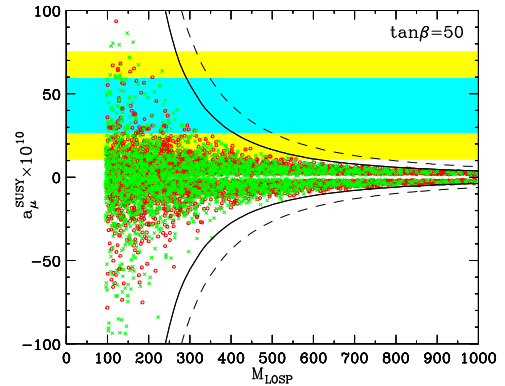}{0.74}
\caption{As in Fig.~\protect\ref{fig:mnlsp}, but assuming a visibly
decaying LSP. }
\label{fig:mlsp}
\end{figure}

In Fig.~\ref{fig:mlsp} we repeat the above analysis, but for the case
where the LSP decays visibly in collider detectors, as in models with
low-scale supersymmetry breaking or $R$-parity violating interactions.
In this case, the LOSP is the LSP.  We relax the requirement of a
neutral LSP, and require slepton masses above 95 GeV and neutralino
masses above 99 GeV~\cite{masslimits}.  The results are given in
Fig.~\ref{fig:mlsp}.  For this case, the solid envelope curve is
parametrized by
\begin{equation}
\frac{\amususy}{43\times 10^{-10}} = 
\frac{\tb}{50} \left[ \! \left( \! 
\frac{300~\gev}{\mlosp^{\text{max}}} \! \right)^2
\!\!\! + \! \left( \! \frac{230~\gev}{\mlosp^{\text{max}}} \!
\right)^4 \right] \, ,
\end{equation}
and the 1$\sigma$ (2$\sigma$) bound is $\mlosp < 410~\gev$ (640 GeV).

These model-independent upper bounds have many implications. They
improve the prospects for observation of weakly-interacting
superpartners at the Tevatron and LHC. They also impact linear
colliders, where the study of supersymmetry requires $\sqrt{s} > 2
\mlosp$ (with the possible exception of associated neutralino
production in stable LSP scenarios).  Finally, these bounds provide
fresh impetus for searches for lepton flavor violation, which is also
mediated by sleptons and charginos/neutralinos.

We now turn to specific models.  The supersymmetric contributions to
$\amu$ have been discussed in various supergravity
theories~\cite{SUGRA}, and more recently in models of
gauge-mediated~\cite{Carena:1997qa,GM} and anomaly-mediated
supersymmetry breaking~\cite{Feng:2000hg,Chattopadhyay:2000ws}.

We first consider the framework of minimal supergravity, in which the
entire weak scale superparticle spectrum is fixed by four continuous
parameters and one binary choice: $m_0$, $\mgaugino$, $A_0$, $\tb$,
and $\sign(\mu)$, where the first three are the universal scalar,
gaugino, and trilinear coupling masses at the grand unified theory
(GUT) scale $\mgut \simeq 2\times 10^{16}~\gev$.  We relate these to
weak scale parameters through two-loop renormalization group
equations~\cite{2loop RGEs} with one-loop threshold corrections and
calculate all superpartner masses to one-loop~\cite{Pierce:1997zz}.
Electroweak symmetry is broken radiatively with a full one-loop
analysis, which determines $|\mu|$.

In minimal supergravity, many potential low-energy effects are
eliminated by scalar degeneracy.  However, $\amususy$ is not
suppressed in this way and may be large.  In this framework,
$\sign(\amususy) = \sign(\mu M_{1,2})$.  As is well-known, however,
the sign of $\mu$ also enters in the supersymmetric contributions to
$\bsg$.  Current constraints on $\bsg$ require $\mu M_3 > 0$ if $\tb$
is large. In minimal supergravity, then, gaugino mass unification
implies that a large discrepancy in $\amu$ is only possible for
$\amususy > 0$, in accord with the new measurement.

\begin{figure}[tbp]
\postscript{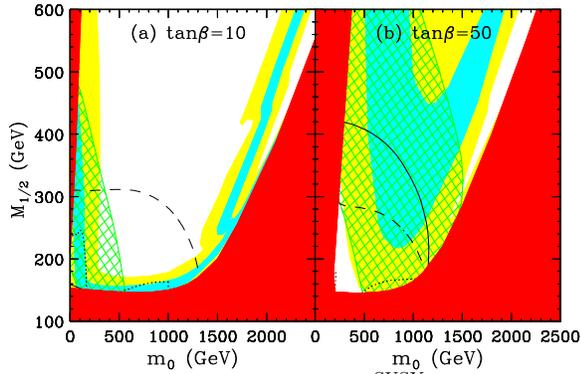}{0.88}
\caption{The 2$\sigma$ allowed region for $\amususy$ (hatched) in
minimal supergravity, for $A_0=0$, $\mu>0$, and two representative
values of $\tb$. The dark red regions are excluded by the requirement
of a neutral LSP and by the chargino mass limit of 103 GeV, and the
medium blue (light yellow) region has LSP relic density $0.1 \le
\Omega h^2 \le 0.3$ ($0.025\le \Omega h^2 \le 1$).  The area below the
solid (dashed) contour is excluded by $\bsg$ (the Higgs boson mass),
and the regions probed by the tri-lepton search at Tevatron Run II are
below the dotted contours.  }
\label{fig:amu_sugra}
\end{figure}

In Fig.~\ref{fig:amu_sugra}, the 2$\sigma$ allowed region for
$\amususy$ is plotted for $\mu > 0$.  Several important constraints
are also included: bounds on the neutralino relic density, the Higgs
boson mass limit $m_h > 113.5~\gev$, and the 2$\sigma$ constraint
$2.18 \times 10^{-4} < B(\bsg) < 4.10 \times 10^{-4}$.

For moderate $\tb$, the region preferred by $\amususy$ is at low
$m_0$.  Much of the favored region is excluded by the Higgs boson
mass. However, the remaining region is consistent with the requirement
of supersymmetric dark matter, and, intriguingly, is roughly that
obtained in no-scale supergravity~\cite{Lahanas:1987uc} and minimal
gaugino-mediated~\cite{Schmaltz:2000gy} models.  In contrast, for
large $\tb$, there is a large allowed area that extends to large
$\mgaugino$ and $m_0 \approx 1.5~\tev$, and which also overlaps
significantly with a region with desirable relic density.  In focus
point models with large and universal scalar
masses~\cite{Feng:2000mn}, large $\tb$ is therefore favored.  The
cosmologically preferred regions of minimal supergravity are probed by
many pre-LHC experiments~\cite{Feng:2000gh}.  Note, however, that the
sign of $\mu$ preferred by $\amu$ implies destructive interference in
the leptonic decays of the second lightest neutralino, and so the
Tevatron search for trileptons is ineffective for $200~\gev < m_0 <
400~\gev$~\cite{Matchev:1999nb}.

We close by considering anomaly-mediated supersymmetry
breaking~\cite{Randall:1999uk}.  One of the most robust and striking
predictions of this framework is that the gaugino masses are
proportional to the corresponding beta function coefficients, and so
$M_{1,2} M_3 < 0$.  Consistency with the $\bsg$ constraint then
implies that only negative $\amususy$ may have large magnitude, in
contrast to the case of conventional supergravity
theories~\cite{Feng:2000hg,Chattopadhyay:2000ws}.

\begin{figure}[tbp]
\postscript{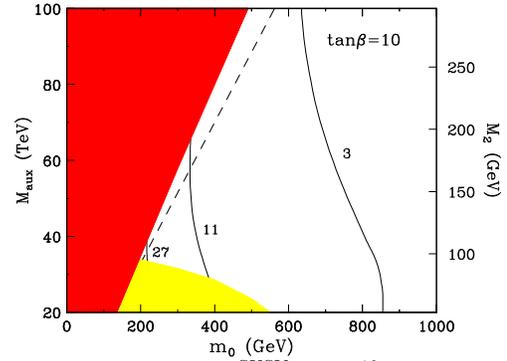}{0.75}
\caption{Contours of $\amususy \times 10^{10}$ in the minimal
anomaly-mediated model, for $\mu>0$ and $\tb = 10$. The dark red
region is excluded by $m_{\tilde{\tau}} > 82~\gev$, the light yellow
region is excluded at 2$\sigma$ by $B(\bsg) < 4.10 \times 10^{-4}$,
and the LSP is a stau to the left of the dashed line. }
\label{fig:amsb}
\end{figure}

In Fig.~\ref{fig:amsb} we investigate how large a positive $\amususy$
may be in the minimal anomaly-mediated model.  This model is
parametrized by $\maux$, $m_0$, $\tb$, and $\sign(\mu)$, where $\maux$
determines the scale of the anomaly-mediated soft terms, and $m_0$ is
a universal scalar mass introduced to remove tachyonic sleptons.  To
get $\amususy >0$, we choose $\mu M_{1,2}> 0$.  We see, however, that
the constraint from $\bsg$ is severe, as this sign of $\mu$ implies a
constructive contribution from charginos to $\bsg$ in anomaly
mediation.  Even allowing a 1$\sigma$ deviation in $\amu$, we have
checked that for all $\tb$, it is barely possible to obtain 2$\sigma$
consistency with the $\bsg$ constraint.  Minimal anomaly mediation is
therefore disfavored. The dependence of this argument on the
characteristic gaugino mass relations of anomaly mediation suggests
that similar conclusions will remain valid beyond the minimal model.

In conclusion, the recently reported deviation in $\amu$ is easily
accommodated in supersymmetric models.  Its value provides {\em
model-independent} upper bounds on masses of observable superpartners
and already discriminates between well-motivated models.  We await the
expected improved measurements with great anticipation.

{\em Acknowledgments} --- This work was supported in part by the
U.~S.~Department of Energy under cooperative research agreement
DF--FC02--94ER40818.

\end{document}